\documentstyle[aps]{revtex}

\begin{document}
\title{Quantum key distribution via quantum encryption}
\author{Yong-Sheng Zhang, Chuan-Feng Li\thanks{%
Electronic address: cfli@ustc.edu.cn}, Guang-Can Guo\thanks{%
Electronic address: gcguo@ustc.edu.cn}}
\address{Laboratory of Quantum Communication and Quantum Computation and Department
of Physics, \\
University of Science and Technology of China, Hefei 230026, P.R. China}
\maketitle

\begin{abstract}
\baselineskip12ptA quantum key distribution protocol based on quantum
encryption is presented in this Brief Report. In this protocol, the
previously shared Einstein-Podolsky-Rosen pairs act as the quantum key to
encode and decode the classical cryptography key. The quantum key is
reusable and the eavesdropper cannot elicit any information from the
particle Alice sends to Bob. The concept of quantum encryption is also
discussed.

PACS number(s): 03.67.Dd, 03.65.Bz\smallskip \vspace{0.3in}
\end{abstract}

\baselineskip12ptThe aim of quantum information science is to use
nonclassical features of quantum systems to achieve performance in
communication and computation that is superior to that achievable with
systems based solely on classical physics. For example, current methods of
public-key cryptography base their security on the supposed (but unproven)
computational difficulty in solving certain problems, such as finding the
prime factors of large numbers - these problems have not only been unproven
to be difficult, but have actually been shown to be computationally ``easy''
in the context of quantum computation \cite{Shor94}. In contrast, it is now
generally accepted that techniques of quantum cryptography can allow
completely secure communications between distant parties \cite{Lo99}. The
problem that the proposed protocols solve is how to enable two protagonists,
``Alice'' and ``Bob,'' who share no secret information initially, to
transmit a secret message $x$, for example, a cryptographic key, under the
nose of an adversary ``Eve,'' who is free to eavesdrop on all their
communications.

In quantum key distribution (QKD), which is one of the important
cryptographic tasks, Alice and Bob's classical communication is supplemented
by a quantum channel, which Eve is also free to eavesdrop on if she dares.
Because of the fragile nature of quantum information, any eavesdropping
disturbs the quantum transmission in a way likely to be detected by Alice
and Bob. The security of protocols for QKD such as the Bennett-Brassard 1984
(BB84) \cite{Ben84} and the Bennett 1992 protocol (B92) \cite{Ben92} is
based on the premise that nonorthogonal states cannot be cloned or
discriminated exactly. Ekert's \cite{Ekt91} and Cabello's \cite{Cab00}
protocols are based on the nonlocal correlation of Einstein-Podolsky-Rosen
(EPR) state. The orthogonal states' quantum QKD protocols are based on
splitting transmission of one bit information into two steps \cite{Gol95}.
We categorize all the protocols mentioned above as source-encrypting QKD,
for their methods are making an alternative choice on the basis of the
source \cite{Ben84,Ben92,Ekt91,Cab00} or splitting the source into two parts 
\cite{Gol95}.

On the other hand, the nonlocal correlation of the EPR \cite{EPR35} state
has been applied to do much work in quantum information field, such as
quantum teleportation \cite{Ben93}, quantum dense coding \cite{Ben922}, QKD 
\cite{Ekt91}, reducing the complexity of communication \cite{Clv97}, etc.
However, other applications of the EPR state in quantum information field
are yet to be discovered.

In this Brief Report, we present a QKD scheme using a method different from
the protocols mentioned above. This protocol is a quantum encryption, i.e.,
using the quantum key to encode and decode the classical information. And
the previously shared reusable EPR state acts as the quantum key. The
information will not be leaked or eavesdropped without being known by the
communication parties. We call it channel-encrypting QKD, compared with the
previous protocols.

The QKD process of this protocol consists of the following steps. Alice and
Bob have previously shared some quantity of the EPR pairs serving as the
quantum key 
\begin{equation}
\left| \Phi ^{+}\right\rangle =\frac 1{\sqrt{2}}\left( \left|
00\right\rangle +\left| 11\right\rangle \right) .  \eqnum{1}
\end{equation}
When the process begins, the two parties rotate their particle's state by
angle $\theta $, respectively. The rotation can be described as 
\begin{equation}
R\left( \theta \right) =\left( 
\begin{array}{cc}
\cos \theta & \sin \theta \\ 
-\sin \theta & \cos \theta
\end{array}
\right) .  \eqnum{2}
\end{equation}
The state $\left| \Phi ^{+}\right\rangle $ does not change under bilateral
operation of $R\left( \theta \right) $. The purpose of this operation is to
prevent the other parties' from eavesdropping. (The detailed interpretation
and the selection of $\theta $ will be given later in this paper.) Then
Alice selects a value of a bit ($0$ or $1$) and prepares a carrier particle $%
\gamma $ in the corresponding state $\left| \psi \right\rangle $ ($\left|
0\right\rangle $ or $\left| 1\right\rangle $) randomly. The classical bit
and the state $\left| \psi \right\rangle $ are only known by Alice herself.
Alice uses the particle $\beta _A$ of the entangled pairs and $\gamma $ in
state $\left| \psi \right\rangle $ to do a controlled-NOT (CNOT) operation ($%
\beta _A$ is the controller and $\gamma $ is the target) and the three
particles will be in a GHZ state 
\begin{eqnarray}
\left| \Psi \right\rangle &=&\frac 1{\sqrt{2}}\left( \left| 000\right\rangle
+\left| 111\right\rangle \right) _{\beta _A\beta _B\gamma }\text{, when }%
\left| \psi \right\rangle =\left| 0\right\rangle ,  \eqnum{3} \\
\text{or }\left| \Psi \right\rangle &=&\frac 1{\sqrt{2}}\left( \left|
001\right\rangle +\left| 110\right\rangle \right) _{\beta _A\beta _B\gamma }%
\text{, when }\left| \psi \right\rangle =\left| 1\right\rangle .  \nonumber
\end{eqnarray}
Then she sends $\gamma $ to Bob. Bob uses his corresponding particle $\beta
_B$ to do a CNOT operation on $\gamma $ again. Now the key particles $\beta
_A$ and $\beta _B$ and the carrier particle $\gamma $ are in the same state
as the initial state 
\begin{equation}
\left| \Psi ^{\prime }\right\rangle =\left| \Phi ^{+}\right\rangle \otimes
\left| \psi \right\rangle .  \eqnum{4}
\end{equation}
Bob measures $\gamma $ and will get the classical bit corresponding to state 
$\left| \psi \right\rangle $.

To assess the secrecy of their communication, Alice and Bob select a random
part of their bit string and compare it over the classical channel.
Obviously, the disclosed bits cannot then be used for encryption anymore. If
their key had been intercepted by an eavesdropper, the correlation between
the values of their bits would have been reduced. Eve's eavesdropping
strategies and the security of this protocol will be discussed later in this
paper.

If the QKD round succeeds, Alice and Bob retain all of the entangled states
and can reuse them the next time. If the round fails, the parties discard
all particles which were used until that point. In this case, Alice and Bob
have to start again with new keys (EPR pairs).

We now discuss the security of this protocol. First, Eve can intercept the
particle Alice sends to Bob and then resend it or another particle to Bob.
However, Eve cannot elicit any information from the particle she
intercepted, because it is in the maximally mixed state 
\begin{equation}
\rho =\frac 12\left( \left| 0\right\rangle \left\langle 0\right| +\left|
1\right\rangle \left\langle 1\right| \right)  \eqnum{5}
\end{equation}
in spite of the bit value Alice sends out. If Eve sends this particle after
disturbing it or sends another particle to Bob, it will introduce error when
Bob decodes it by using the quantum key. If the state of the particle sent
by Alice has been changed, the final state assumed is 
\begin{eqnarray}
\rho _i &=&\sum_{k=1}^2p_{ik}\left| \Psi _{ik}^{\prime }\right\rangle
\left\langle \Psi _{ik}^{\prime }\right| ,  \eqnum{6} \\
\left| \Psi _{ik}^{\prime }\right\rangle &=&a_{ik}^{\prime }\left|
0\right\rangle +b_{ik}^{\prime }\left| 1\right\rangle .  \nonumber
\end{eqnarray}
The average error rate of the classical key Alice transmits to Bob will be $%
\frac 12$.

The second eavesdropping strategy is to entangle with the key. Eve can
intercept the particle $\gamma $ Alice sends to Bob and use it and her own
particle in state $\left| 0\right\rangle $ or $\left| 1\right\rangle $ to do
a CNOT operation (her own particle is the target and $\gamma $ is the
controller). Then Eve resends $\gamma $ to Bob. After Bob's decoding
operation, Eve's particle is entangled with the key. It seems that Eve can
use her particle to decode Alice's particle next time as Bob does. However,
Eve cannot know in which state she is entangled with the key and cannot get
any information of the state Alice sends to Bob. To detect this
eavesdropping strategy, Alice and Bob can do a bilateral rotation $R(\theta
) $ on the key (EPR\ pairs in state $\left| \Phi ^{+}\right\rangle $) before
Alice does the CNOT operation. The state of the maximally entangled two
particles will be unchanged in this case. If Eve has entangled her particle
with Alice and Bob's particles in the state $\left| \Phi \right\rangle
_{ABE}=\frac 1{\sqrt{2}}\left( \left| 000\right\rangle +\left|
111\right\rangle \right) $ [or $\left| \Phi \right\rangle _{ABE}=\frac 1{%
\sqrt{2}}\left( \left| 001\right\rangle +\left| 110\right\rangle \right) $].
In the second round, the entangled state will be changed to 
\begin{eqnarray}
\left| \Phi \right\rangle _{ABE} &=&\cos ^2\theta \frac 1{\sqrt{2}}\left(
\left| 000\right\rangle +\left| 111\right\rangle \right) +\sin ^2\theta 
\frac 1{\sqrt{2}}\left( \left| 110\right\rangle +\left| 001\right\rangle
\right)  \eqnum{7} \\
&&\ +\sin \theta \cos \theta \frac 1{\sqrt{2}}\left( \left| 011\right\rangle
-\left| 100\right\rangle \right) +\sin \theta \cos \theta \frac 1{\sqrt{2}}%
\left( \left| 101\right\rangle -\left| 010\right\rangle \right)  \nonumber
\end{eqnarray}
or 
\begin{eqnarray}
\left| \Phi \right\rangle _{ABE} &=&\cos ^2\theta \frac 1{\sqrt{2}}\left(
\left| 001\right\rangle +\left| 110\right\rangle \right) +\sin ^2\theta 
\frac 1{\sqrt{2}}\left( \left| 111\right\rangle +\left| 000\right\rangle
\right)  \eqnum{8} \\
&&\ +\sin \theta \cos \theta \frac 1{\sqrt{2}}\left( \left| 010\right\rangle
-\left| 101\right\rangle \right) +\sin \theta \cos \theta \frac 1{\sqrt{2}}%
\left( \left| 100\right\rangle -\left| 011\right\rangle \right)  \nonumber
\end{eqnarray}
under the bilateral rotation. The error rate of the bit that Alice sends to
Bob will be $2\cos ^2\theta \sin ^2\theta $. So if $\theta =\frac \pi 4$,
the error rate the eavesdropping caused will reach $\frac 12$. Thus the
communication parties can select $\theta =\frac \pi 4$ as the bilateral
rotation angle in every round and Eve cannot get any useful information of
the bit string they transmit.

Then we consider the more generic attacking. Assume that Eve can use her own
system and the qubits sent by Alice to do a completely positive trace
preserving map 
\begin{equation}
\Lambda \left( \rho \right) =\sum_iV_i\rho V_i^{\dagger }\text{ with }%
\sum_iV_iV_i^{\dagger }=I  \eqnum{9}
\end{equation}
on them, where $I$ denotes the identity operator on the Hilbert space of the
whole system's state. It is known \cite{Kraus} that the map is also of the
form 
\begin{equation}
\Lambda \left( \rho \right) =Tr_C\left[ U\rho \otimes \omega U^{\dagger
}\right] ,  \eqnum{10}
\end{equation}
where $\omega $ is a state on the additional system $C$, and $U$ is the
unitary transformation on the joint system. So Eve's completely positive
trace preserving map is equal to a unitary transformation on a larger
system, and we can only consider the case in which Eve tries to obtain the
information by unitary transformation on her own entire system and the qubit
sent by Alice.

Suppose that Eve's system has entangled with Alice and Bob's key in the
state 
\begin{equation}
\frac 1{\sqrt{2}}\left( \left| 0\right\rangle \left| 0\right\rangle \left|
\psi _0\right\rangle +\left| 1\right\rangle \left| 1\right\rangle \left|
\psi _1\right\rangle \right) _{ABE}\otimes \left| 0\right\rangle _I, 
\eqnum{11}
\end{equation}
where $\left| 0\right\rangle _I$ is an indicator qubit for Eve to detect
Alice's sending qubit and there is no restriction on the form of $\left|
\psi _0\right\rangle $ and $\left| \psi _1\right\rangle $. After Alice and
Bob do a bilateral rotation $R\left( \theta \right) $, Alice does a CNOT
operation on the sending qubit and sends it out. Then Eve does a unitary
transformation on the sending qubit and her own system. She expects that the
indicator will be the same as the sending qubit. Assume that the unitary
transformation has the universal form 
\begin{eqnarray}
U\left| i\right\rangle _S\left| \psi _k\right\rangle _E\left| j\right\rangle
_I &=&(a_{ijk}\left| 0\right\rangle \left| \psi _{aijk}\right\rangle \left|
0\right\rangle +b_{ijk}\left| 0\right\rangle \left| \psi
_{bijk}\right\rangle \left| 1\right\rangle  \eqnum{12} \\
&&\ +c_{ijk}\left| 1\right\rangle \left| \psi _{cijk}\right\rangle \left|
0\right\rangle +d_{ijk}\left| 1\right\rangle \left| \psi
_{dijk}\right\rangle \left| 1\right\rangle )_{SEI},  \nonumber
\end{eqnarray}
where $i,j,k=0,1$ and there is no restriction on the final states of $\left|
\psi \right\rangle _E$.

If the attack strategy is successful, it requires that the output of the
indictor qubit be the same as the sending qubit. However, when we compute
the unitary transformation in Eq. (12) to satisfy this condition, we find
that all the factors in Eq. (12) must be equal to $0$. It means that the
unitary transformation for a successful attacking strategy does not exist
and the protocol is secure under this type of attack strategy. In fact,
Eve's system and the sending qubit are in the reduced state 
\begin{equation}
\rho =\frac 14\left( \left| 0\right\rangle \left\langle 0\right| +\left|
1\right\rangle \left\langle 1\right| \right) _S\otimes \left( \left| \psi
_0\right\rangle \left\langle \psi _0\right| +\left| \psi _1\right\rangle
\left\langle \psi _1\right| \right) _E\otimes \left| 0\right\rangle
_I\left\langle 0\right| ,  \eqnum{13}
\end{equation}
whether the qubit sent by Alice is $\left| 0\right\rangle $ or $\left|
1\right\rangle $. So Eve has no way to distinguish them and to obtain the
qubit sent by Alice. Even if Eve adopts stronger strategies that would cause
fewer errors and thus might be able to hide her presence in channel noise,
she will obtain no information of the classical bit Alice sends to Bob,
though she cannot be found.

Up to this point, our discussion has assumed that the initial state is the
ideal maximally entangled state $\left| \Phi ^{+}\right\rangle $. Suppose,
however, that this state is corrupted a little after it is reused for many
times, due to nonexact operation or decoherence. Alice and Bob have a state
described by the density matrix 
\begin{equation}
\rho =\left( 1-\epsilon \right) \left| \Phi ^{+}\right\rangle \left\langle
\Phi ^{+}\right| +\epsilon \rho _1,  \eqnum{14}
\end{equation}
where $\epsilon $ is a parameter of the deviation of $\rho $ from $\left|
\Phi ^{+}\right\rangle \left\langle \Phi ^{+}\right| $ and $\rho _1$ is
density matrix of an arbitrary state. Our results are most easily presented
using the {\it trace distance}, a metric on Hermitian operators defined by $%
T\left( A,B\right) =Tr\left( \left| A-B\right| \right) $ \cite{Vi99}, where $%
\left| X\right| $ denotes the positive square root of the Hermitian matrix $%
X^2$. From the above, we can get that $T\left( \left| \Phi ^{+}\right\rangle
\left\langle \Phi ^{+}\right| ,\rho \right) \leq 2\sqrt{\epsilon }$.

Ruskai \cite{Ru94} has shown that the {\it trace distance} contracts under
physical processes. If all operations are exact in the next QKD process,
since the state $\left| \Phi ^{+}\right\rangle $ will be unchanged in the
process, the density matrix $\rho $ will be transformed to 
\begin{equation}
\rho ^{\prime }=\left( 1-\epsilon \right) \left| \Phi ^{+}\right\rangle
\left\langle \Phi ^{+}\right| +\epsilon \rho _1^{\prime },  \eqnum{15}
\end{equation}
and $T\left( \left| \Phi ^{+}\right\rangle \left\langle \Phi ^{+}\right|
,\rho ^{\prime }\right) \leq 2\sqrt{\epsilon }$.

The fidelities \cite{Joz94} $F\left( \left| \Phi ^{+}\right\rangle
\left\langle \Phi ^{+}\right| ,\rho \right) $ and $F\left( \left| \Phi
^{+}\right\rangle \left\langle \Phi ^{+}\right| ,\rho ^{\prime }\right) $
are both no less than $1-\epsilon $, so the probability that the QKD process
fails is no more than $\epsilon $. Therefore, we can say that this protocol
is robust. To prevent the degeneration of the entangled state, technologies
of quantum privacy amplification \cite{Deu96} and entanglement purification 
\cite{Ben96} can be used. These processes need only local quantum operation
and classical communication (LQCC). However, the number of available
entangled pairs will be reduced and need to be supplied by sending qubits.

In the case where Alice sends particles in a noisy channel, the channel can
be described by the Kraus operator \cite{Kraus} 
\begin{equation}
\rho ^{\prime }=\sum_\mu M_\mu ^{\dagger }\rho M_\mu ,  \eqnum{16}
\end{equation}
\begin{equation}
M_0=\sqrt{1-p_1-p_2-p_3}I,M_1=\sqrt{p_1}\sigma _1,M_2=\sqrt{p_2}\sigma
_2,M_3=\sqrt{p_3}\sigma _3,  \eqnum{17}
\end{equation}
where $I$ is the identity operator and $\sigma _i$ are Pauli operators 
\begin{equation}
I=\left( 
\begin{array}{cc}
1 & 0 \\ 
0 & 1
\end{array}
\right) ,\sigma _1=\left( 
\begin{array}{cc}
0 & 1 \\ 
1 & 0
\end{array}
\right) ,\sigma _2=\left( 
\begin{array}{cc}
0 & -1 \\ 
1 & 0
\end{array}
\right) ,\sigma _3=\left( 
\begin{array}{cc}
1 & 0 \\ 
0 & -1
\end{array}
\right) .  \eqnum{19}
\end{equation}
Since the original states Alice wants to send are orthogonal states in the
basis $\left\{ \left| 0\right\rangle ,\left| 1\right\rangle \right\} $,
after Bob's decoding and measurement, the error of the state will be
projected to one of the three Pauli operators $\sigma _i$, with probability $%
p_i$, respectively. For error $\sigma _1$, the carrier bit is flipped and
the EPR pair is not affected, for $\sigma _3$, the carrier bit is not
affected but the state of the EPR pair is changed from $\left| \Phi
^{+}\right\rangle $ to $\left| \Phi ^{-}\right\rangle =\frac 1{\sqrt{2}}%
\left( \left| 00\right\rangle -\left| 11\right\rangle \right) $, and for $%
\sigma _2$, the carrier bit is flipped and the EPR state is changed to $%
\left| \Phi ^{-}\right\rangle $. As discussed later in this paper, we can
conquer the bit flip by duplication code, And the corruption of the EPR
pairs is the same as mentioned above.

In this protocol, the previously shared EPR pairs act as a quantum key to
encode and decode the classical cryptography key, and the quantum key is
reusable. In classical cryptography, both the encryption key and the
decoding key are random series, but they have a definite correlation. It is
the randomness in the encryption key that makes the information secure and
it is the correlation between the decoding key and the encryption key that
makes the receiver able to extract the useful information from the
cryptogram. Similar to the classical counterpart, in the quantum encryption
we presented, from any party's point of view the quantum key is in a
maximally mixed state. In fact, any single particle of the entangled pair is
in a completely uncertain state. However, the states of the two parties'
particles have strong quantum correlation, which is called entanglement, and
have no classical counterpart. It is this correlation that makes the quantum
encryption secure. Since this correlation cannot be produced by LQCC, the
eavesdropper cannot establish this correlation with the sender. So the
quantum key is reusable. In the classical case, the only crypto system that
provides perfect secrecy is the ``one time pad'' system, in which the
encryption key cannot be used repeatedly. The more interesting character of
this QKD scheme is that, since the eavesdropper cannot elicit any
information from the particle Alice sends to Bob, Alice can use classical
error-correction code technology, such as duplication code, to conquer the
bit flip error.

From another point of view, this protocol can be regarded as a quantum
channel encryption, i.e., a quantum key encrypts the quantum channel. In the
QKD protocols presented before, the sender uses alternative choices of the
basis of the source or sends out the qubit separately. In our protocol, the
classical bit is represented by the normal orthogonal basis of the particle.
If we regard the EPR pair as a part of the channel, the quantum channel is
encrypted. The classical information source can be transmitted securely,
without leaking, in this modified channel.

In practice, the previously shared EPR pairs can be realized by standing
qubits, such as entangled atoms, and the single photons can serve as the
flying qubits sent out by Alice. The interaction of atom and photon can be
realized by the technology of cavity quantum electrodynamics (CQED). The
theoretical schemes of this technology have been proposed \cite{Slt95} and
the research in laboratory has made some progress \cite{QED}. So this
protocol is expected to be realized in the laboratory in the near future.

In summary, we have presented the concept of quantum encryption and proposed
a QKD scheme based on quantum encryption. This method has been used in
quantum authentication \cite{Zha00} and can be used for the encryption of
arbitrary quantum states \cite{Leu}.

This work was supported by the National Natural Science Foundation of China.

\end{document}